# Urban Planning in 3D with a Two-tier LUTI model

Flora Roumpani[1,2], Joel Dearden[1†], Alan Wilson[1,2]

## ABSTRACT

The two-tier Lowry model brings dynamic simulations of population and employment directly into the planning process. By linking regional modelling with neighbourhood design, the framework enables planners to explore how alternative planning scenarios may evolve over time. The upper tier captures regional flows of people, jobs, and services, while the lower tier allocates these to fine-grain zones such as neighbourhoods or parcels. Implemented in CityEngine, the approach allows interactive visualisation and evaluation of multi-scale scenarios. A case study in South Yorkshire (UK) illustrates how regional forecasts can be translated into local design responses, connecting quantitative modelling with 3D spatial planning.

## 1. Introduction

A typical urban planning task involves developing scenarios and testing different allocations of housing, social infrastructure, and other facilities across multiple land-use zones. By contrast, urban modelling and regional planning generally work on a single zonal system to calculate flows between origins and destinations. Models such as Land Use–Transport Interaction (LUTI) have been successful at the regional scale, but their integration into design workflows has been limited, constrained by computational complexity, high data requirements, and disciplinary separation between modelling and design practice.

This disconnect means that planners often lack tools that link broad-scale dynamics with local-scale decisions. For example, population growth predicted at the regional level may drive local demand for housing, schools, and services, yet such relationships are rarely incorporated into the early stages of design. Addressing this requires methods that can capture the interdependencies between scales and

---

[1] The Alan Turing Institute, London, United Kingdom
[2] Centre for Advanced Spatial Analysis, University College London, United Kingdom
† *In memoriam (1978–2020)*

Corresponding Author: f.roumpani@turing.ac.uk

provide a synergetic framework where regional forecasts and neighbourhood design inform one another.

Previous and current attempts to bridge this divide include systems such as UrbanSim (Waddell, 2002), QUANT (Batty, 2021; Batty & Milton, 2023), and emerging urban digital twins. While powerful, these approaches are often data-intensive and computationally heavy, making them less suitable for early-stage design. Our approach builds on earlier work on procedural urban modelling as an interactive tool for planning (Roumpani, 2013; 2023), combining the rigour of a Lowry-type LUTI model with direct integration into a 3D design environment. This lighter, design-oriented framework bridges urban modelling and masterplanning, enabling planners to evaluate scenarios within a single visual platform.

Although this paper demonstrates the approach with a Dynamic Lowry Model for South Yorkshire in UK, the framework is designed to be modular in principle, so that other regional models could be substituted as the upper tier if desired. We use this implementation to test how regional forecasts of population and employment can be translated into local planning data to evaluate housing demand, service provision, and alternative layouts.

## 2. The Two-Tier model

Assume there are two nested zone systems: an **upper tier** (I, J, K …), representing regions or wards, and a **lower tier** (i, j, k …), representing finer zones such as neighbourhoods, land-use zones, or blocks. For simplicity, we assume that lower-tier zones aggregate into upper-tier ones, so that i ∈ J means lower-tier zone *i* is contained in upper-tier zone *J*.

The purpose of the two-tier structure is to calculate flows T (e.g., commuting trips, shopping trips, school travel) across both systems. This creates multiple possible interactions:

$S_{IJ}$: *upper-tier zone to upper-tier zone*

$T_{ij}$: *lower-tier zone to lower-tier zone*

$U_{iJ}$ or $T_{Ij(J)}$: *lower-tier to upper-tier*

$V_{Ij}$ or $T_{Jj(I)}$: *upper-tier zone to a lower-tier zone*

To develop the two-tier structure, we use a doubly constrained Spatial Interaction Model (SIM) derived from entropy maximisation (Wilson, 1967; 1970) and widely presented in transport modelling texts (Ortúzar & Willumsen, 2011; 2021). A singly constrained SIM is typically used when flows are fixed only at one end, for example in retail modelling where the total demand is known but destinations compete for it. In contrast, the doubly constrained form fixed flows at both origins and destinations, making it suitable for cases such as commuting, where both the number of residents and the number of jobs must be satisfied. The original equations of the doubly constrained model are:

$$T_{ij} = A_i O_i B_j D_j f(c_{ij}) \qquad (1)$$

where $O_i$ and $D_j$ are the origin and destination totals, $f(c_{ij})$ is a deterrence function of travel cost, and $A_i$ and $B_j$ are balancing factors. These factors are updated iteratively using the Furness method (Furness, 1965):

$$A_i = \frac{1}{\Sigma_j B_j D_j f(c_{ij})} \qquad (2)$$

$$B_j = \frac{1}{\Sigma_i A_i O_i f(c_{ij})} \qquad (3)$$

Iterations proceed until convergence, defined as the point where the change in balancing factors between successive steps falls below a pre-specified threshold p:

$$p = |A_i(x-1) - A_i(x)| \qquad (4)$$

Suppose now we want to test a new housing development in an upper-tier zone J, while allocating new schools in lower-tier zones j ∈ K. We can adjust the model as follows:

1. Upper-tier flows ($T_{IJ}$) estimate movements between regions.
2. Link arrays ($U_{iJ}$, $V_{Jj}$) capture how flows cross scales between the two systems.
3. Lower-tier flows ($T_{ij}$) distribute population and service demand within the neighbourhood system.

To adjust push–pull dynamics between scales, we define scaling factors f,g:

$$f_I = \frac{E_I S_{IK}}{S_I^*}, \qquad g_J = \frac{H_J S_{IK}}{S_J^*} \qquad (5\text{-}6)$$

where E and H represent employment and housing, and the asterisk denotes summation over the relevant index. For convenience, we define $S_{IK}=\sum_{j \in K} T_{Ij}$ as the total flow from upper-tier zone I into the lower-tier system K, and $S_{KJ}=\sum_{i \in K} T_{iJ}$ as the total flow from K into zone J. These definitions allow us to write subsequent constraints more compactly.

The constraints that have to be satisfied are:

$$\sum_{j \in K} T_{ij} + \sum_{J \neq K} U_{iJ} = e_i, \quad i \in K \qquad (7)$$

$$\sum_{i \in K} T_{ij} + \sum_{I \neq K} V_{Ij} = h_j, \quad j \in K \quad (8)$$

$$\sum_i U_{iJ} = S_{KJ}, \quad J \neq K \quad (9)$$

$$\sum_j V_{Ij} = S_{IK}, \quad I \neq K \quad (10)$$

Introducing balancing factors for each constraint ($A_i$, $B_j$, $D_J$, $C_I$), the model can be written as:

$$T_{ij} = A_i B_j e_i h_j \exp(-\beta c_{ij}), \quad i, j \in K \quad (10)$$

$$U_{iJ} = A_i D_J e_i H_J \exp(-\beta c_{iJ}), \quad i \in K, J \neq K \quad (11)$$

$$V_{Ji} = B_j C_I E_I h_j \exp(-\beta c_{iJ}), \quad j \in K, I \neq K \quad (12)$$

where c represents cost (e.g. distance or travel time) and β a cost-decay parameter.

Substituting Eqs. (11)–(13) into the constraints (7)–(10) gives:

$$\sum_{j \in K} A_i B_j e_i h_j \exp(-\beta c_{ij}) + \sum_{I \neq K} A_i D_J e_i H_J \exp(-\beta c_{iJ}) = e_i, \quad i \in K \quad (13)$$

$$\sum_{i \in K} A_i B_j e_i h_j \exp(-\beta c_{ij}) + \sum_{I \neq K} B_j C_I E_I h_j \exp(-\beta c_{Ij}) = h_j, \quad j \in K \quad (14)$$

$$\sum_{i \in K} A_i D_J e_i H_J \exp(-\beta c_{iJ}) = S_{KJ}, \quad J \neq K \quad (15)$$

$$\sum_{j \in K} B_i C_I E_I h_i \exp(-\beta c_{Ij}) = S_{IK}, \quad I \neq K \quad (16)$$

From these, the balancing factors follow:

$$A_i = \frac{1}{\sum_{j \in K} B_j h_j \exp(-\beta c_{ij}) + \sum_{J \neq K} D_J H_J \exp(-\beta c_{iJ})}, \quad i \in K \quad (17)$$

$$B_j = \frac{1}{\sum_{i \in K} A_i e_i \exp(-\beta c_{ij}) + \sum_{I \neq K} C_I E_I \exp(-\beta c_{Ij})}, \quad j \in K \quad (18)$$

$$D_J = \frac{S_{KJ}}{H_J \sum_{i \in K} A_i e_i \exp(-\beta c_{iJ})}, \quad J \neq K \quad (19)$$

$$C_I = \frac{S_{IK}}{E_I \sum_{j \in K} B_j h_j \exp(-\beta c_{Ij})}, \quad I \neq K \quad (20)$$

We have used the relationships in (5) and (6) to simplify equations (20) and (21).

It should be then possible to solve these iteratively starting with $C_I = D_J = B_j = 1$, and then iterate with:

1. update A$_i$ via (18); update B$_j$ via (19);
2. update D$_J$ via (20); update C$_I$ via (21).

and cycle until convergence, i.e., when the iteration change falls below threshold p (cf. (4)).

This scheme enforces the origin/destination totals at both tiers while consistently injecting external flows S$_{KJ}$ and S$_{IK}$ from the upper tier into the lower-tier allocation, exactly what we need to link regional pressures to neighbourhood design.

## 3. Cases – Variations of the Two-Tier

As Alexander reminds us in *A City is Not a Tree*, cities are seldom neatly nested structures: their zones often overlap or cut across administrative boundaries. For planners, this means that the choice of lower-tier system depends on the problem at hand. Population allocation to housing may require a finer zonal system than retail modelling, and service catchments (e.g., schools) may extend across multiple regional units. Moreover, detailed local data are not always available.

Below we outline three practical cases that may arise in real-world applications and show how the two-tier model can accommodate each:

1. Case 1: Detailed local data (e$_i$, h$_j$) available and zonal boundaries are consistent.
2. Case 2: Lower-tier data (e$_i$, h$_j$) unknown or partial, but zonal boundaries are consistent.
3. Case 3: Catchments cross K's boundaries, requiring expansion.

**3.1 Case 1:** Data-rich neighborhoods with regional consistency

Assume we have data on employment and housing for both the upper-tier system (large zones such as districts: I,J = A,B..) and the lower-tier zones (smaller zones: i,j = 1,2,3.. ∈ K). The goal is to model commuting flows inside the zone K (T$_{ij}$), as well as flows in and outside K (T$_{iJ}$, T$_{Ij}$).

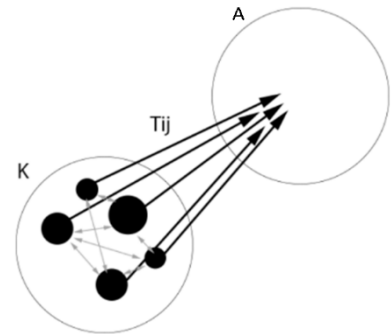

Figure 1. The lower-tier model estimates all flows in and out of zone K, ensuring consistency with the upper-tier totals T$_{KJ}$ and T$_{IK}$.

In this case, the upper tier flows to and out of zone K can be handled by the lower tier model by treating the regional zones, as lower tier subzones (A$_I$=A$_i$, A$_J$=A$_j$). The requirement is then that the detailed lower-tier flows must add up to the totals given by the upper-tier model:

$$\sum_{i \varepsilon K} T_{iJ} = T_{KJ} \quad (22)$$

$$\sum_{j \varepsilon K} T_{Ij} = T_{IK} \quad (23)$$

Balancing factors are introduced to the model, to enforce these equalities, just like in a doubly-constrained gravity model.

**3.2 Case 2:** Data-poor neighborhoods with proportional allocation

Imagine 1,000 people commute from zone A into zone K, but it is unknown which neighbourhoods inside K they go to. If we know each neighbourhood's share of housing, we can split the commuters proportionally. The result may not be as precise as if detailed survey data were available, but it remains consistent with the regional total.

Thus, when there are no reliable lower-tier data ($e_i$, $h_j$), but the total flows at the regional level are known ($T_{IK}, T_{KJ}$), we can distribute the totals into the subzones using weights:

- From I→K: split the total $T_{IK}$ into $\{T_{Ij}\}$ according to weights (e.g., capacities of housing $h_j$).

- From K→J: split the total $T_{KJ}$ into $\{T_{iJ}\}$ according to weights (e.g., proportion of jobs $e_i$).

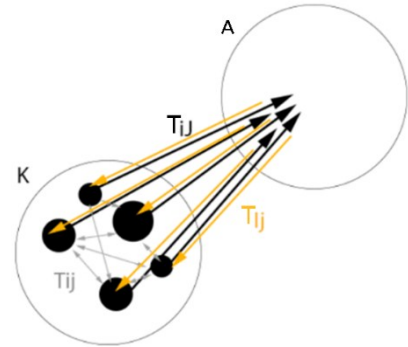

Figure 2. Regional totals ($T_{IK}$, $T_{KJ}$) are split into lower-tier flows $\{T_{Ij}\},\{T_{iJ}\}$ according to weights such as housing or employment capacities.

This way, the detailed flows are consistent "by construction", because they always add up to the upper-tier totals.

**3.3 Case 3:** Catchments that cross boundaries

Imagine you are planning new schools in zone K, that serves children from both inside and outside K. Or a hospital catchment spills over into the next district. That means the detailed flows cannot be modelled correctly unless the system is expanded.

To model such cases, K can be expanded by converting adjacent upper-tier zones into additional lower-tier zones. This creates a buffer of fine-grain zones around the area of interest, allowing cross-boundary flows to be captured explicitly. However, this introduces potential boundary inconsistencies and risks of model misspecification, issues linked to the *modifiable areal unit problem* (Openshaw, 1984; Upton & Fingleton, 1985; Fotheringham & Wong, 1991).

A practical solution is to disaggregate regional indicators (e.g., employment, housing) into the new subzones in proportion to area or land-use shares (Flowerdew & Green, 1994). After this adjustment, the expanded system can be treated as either Case 1 or Case 2, depending on whether detailed local indicators are available.

In planning terms, this corresponds to extending a study area to include adjacent neighbourhoods when catchments — for example, for schools or hospitals — extend beyond administrative boundaries. Regional data are "borrowed" from neighbouring areas and reassigned to finer subzones, ensuring that both local and regional totals remain consistent.

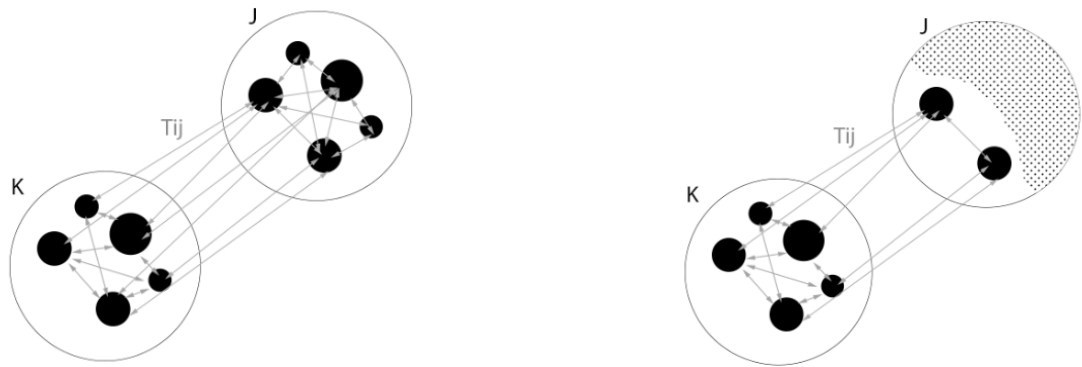

Figure 3. Examples of expanded study areas for cross-boundary catchments. The diagrams shows how adjacent upper-tier zones can be converted into lower-tier subdivisions to capture flows between neighbouring districts.

## 4. A Two-Tier Structure in South Yorkshire

We apply the two-tier model in a planning context using South Yorkshire (UK), a county of ~1.5 million residents and ~3,900 km². The area has been the focus of long-term regeneration and ERDF support, with multiple planned developments (e.g., in Doncaster). We use these conditions to demonstrate two coupled model runs:

1. Upper tier (regional): An inter-/intra-urban model based on the Dynamic Lowry Model (DLM) of Dearden & Wilson (2015).

2. Lower tier (planning scale): A simple, planner-facing model for neighbourhood allocation of residences, schools, and services.

Together, these runs show how regional dynamics can be translated into 3D design scenarios for a specific study area K (here, within Doncaster).

### 4.1 The South Yorkshire Dynamic Lowry Model

The DLM forecasts land-use patterns and population flows across South Yorkshire. It is a Lowry-type regional model that (i) allocates population and services from basic employment, and (ii) uses a calibrated data pipeline (microsimulation outputs) to reproduce observed conditions. The model contains on the order of ~100 variables and simulates decades of change (e.g., 20–50 years) via iterative updates.

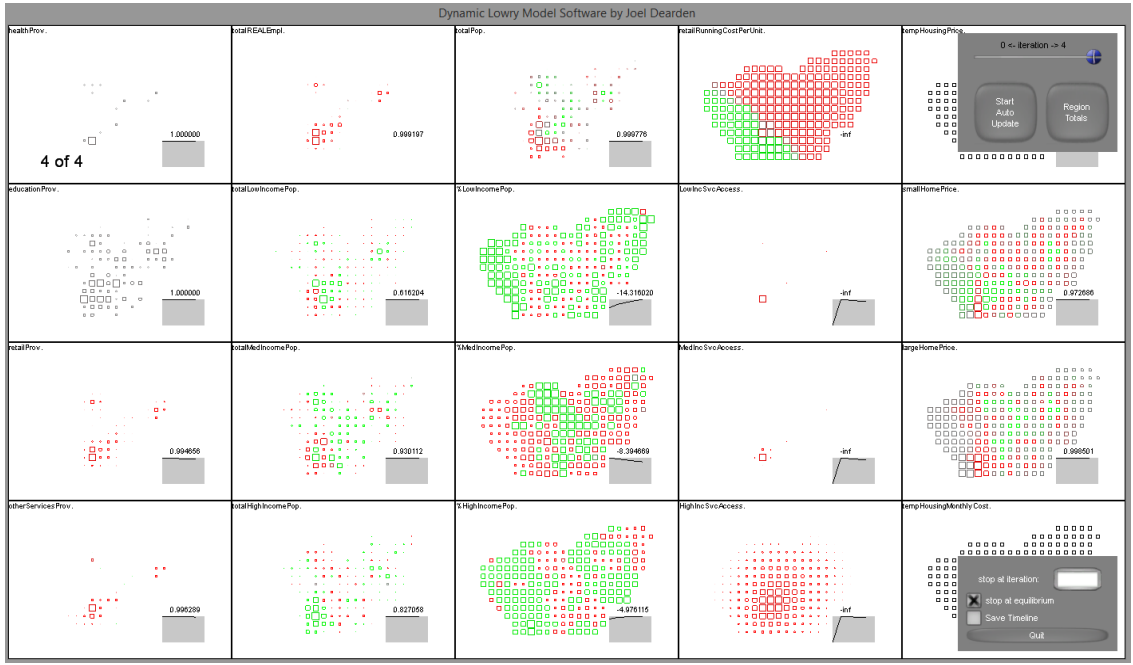

Figure 4. The Dynamic Lowry model for South Yorkshire developed by J. Dearden (2015). Each panel presents forecasts for a different model indicator, allowing comparison of population, employment, and service dynamics across the region.

Land uses are represented for industry, residential, and services, with services distinguished as consumer-driven (e.g., retail) and regulated (e.g., health, education). Residential demand is disaggregated by income classes; housing supply by type (e.g., large/small/temporary). Density and minimum-size constraints ensure feasible allocations.

The DLM's calibrated baseline and sector detail make it a robust upper-tier source of external flows ($T_{IK}$,$T_{KJ}$) and provides indicators (population, employment, service provision) to drive our lower-tier planning model. Further details and limitations are described in Dearden and Wilson (2015).

Although the DLM runs as a standalone application (Figure 4), it provides zone-level outputs per iteration that can be streamed into CityEngine. The regional zoning is a 9 km² grid of 211 labelled cells. Each iteration updates indicators and growth rates that are passed to the two-tier system to calculate external flows ($S_{KJ}$ and $S_{IK}$).

### 4.2 A Two-Tier System in CityEngine

To couple models and design we implement an Information System (IS) in CityEngine and Python (Figure 5). CityEngine, is a procedural urban modelling environment originally developed by Pascal Müller and colleagues (Müller et al., 2006), which enables the rule-based generation and visualization of 3D urban form. The IS will import the DLM in CityEngine and will run a second internal model to translate inputs to planning zones. The IS:

1. **Ingests data** – static GIS layers (parcels, land use, building heights), planning datasets (Unitary Development Plan), and dynamic DLM outputs.

2. **Performs model coupling** – passing DLM totals and indicators ($T_{IK}, T_{KJ}$), sectoral measures) into the lower tier, where a local allocation model runs over fine zones in K.

3. **Provides interaction** – planner-facing controls (sliders, toggles) for switching indicators, adjusting assumptions, and editing 3D layouts.

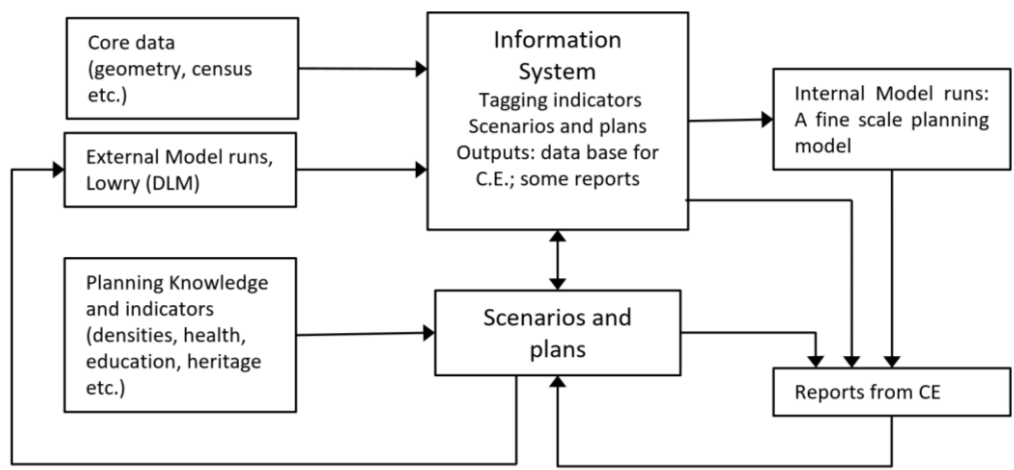

Figure 5. An interactive system of data and model and planning outputs in CityEngine

### 4.3 Upper tier: Dynamic Lowry Model (DLM) in CityEngine

The 211 DLM zones are imported into CityEngine as labelled polygons with attribute data (population, employment, services). Parameters are exposed via a small UI (Figure 7), with exogenous inputs (basic employment, attractiveness factors, area per employee/resident), baseline outputs (population, employment, services), and iterative outputs (rents, land use, service provision, classed populations, housing prices). Indicators can be visualised as 3D bars or thematic overlays, with bar height showing magnitude and colour/time-series animation showing growth or decline. Users can also modify exogenous variables and feed them back into the DLM to generate new scenarios. These controls enable quick scenario toggling, turning CityEngine into a "visualization machine" for the DLM.

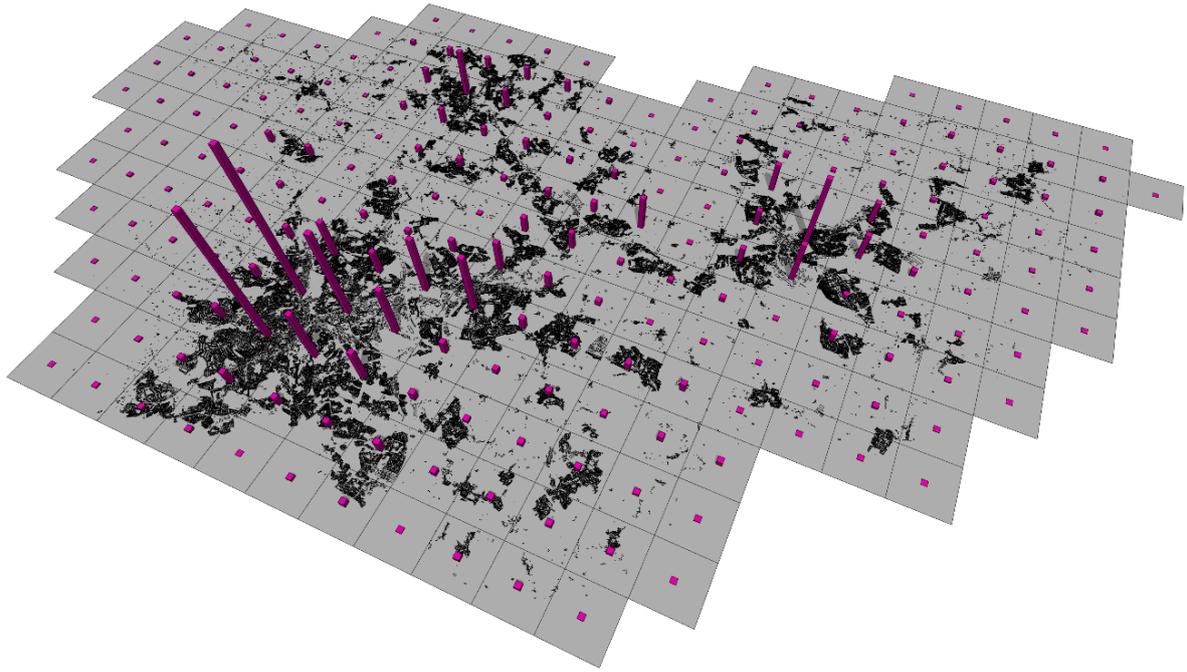

Figure 6. Two-tier implementation in CityEngine. The image shows the 211-cell DLM grid (upper tier) with employment baselines represented as 3D bars alongside fine-grained planning zones from the Doncaster masterplan.

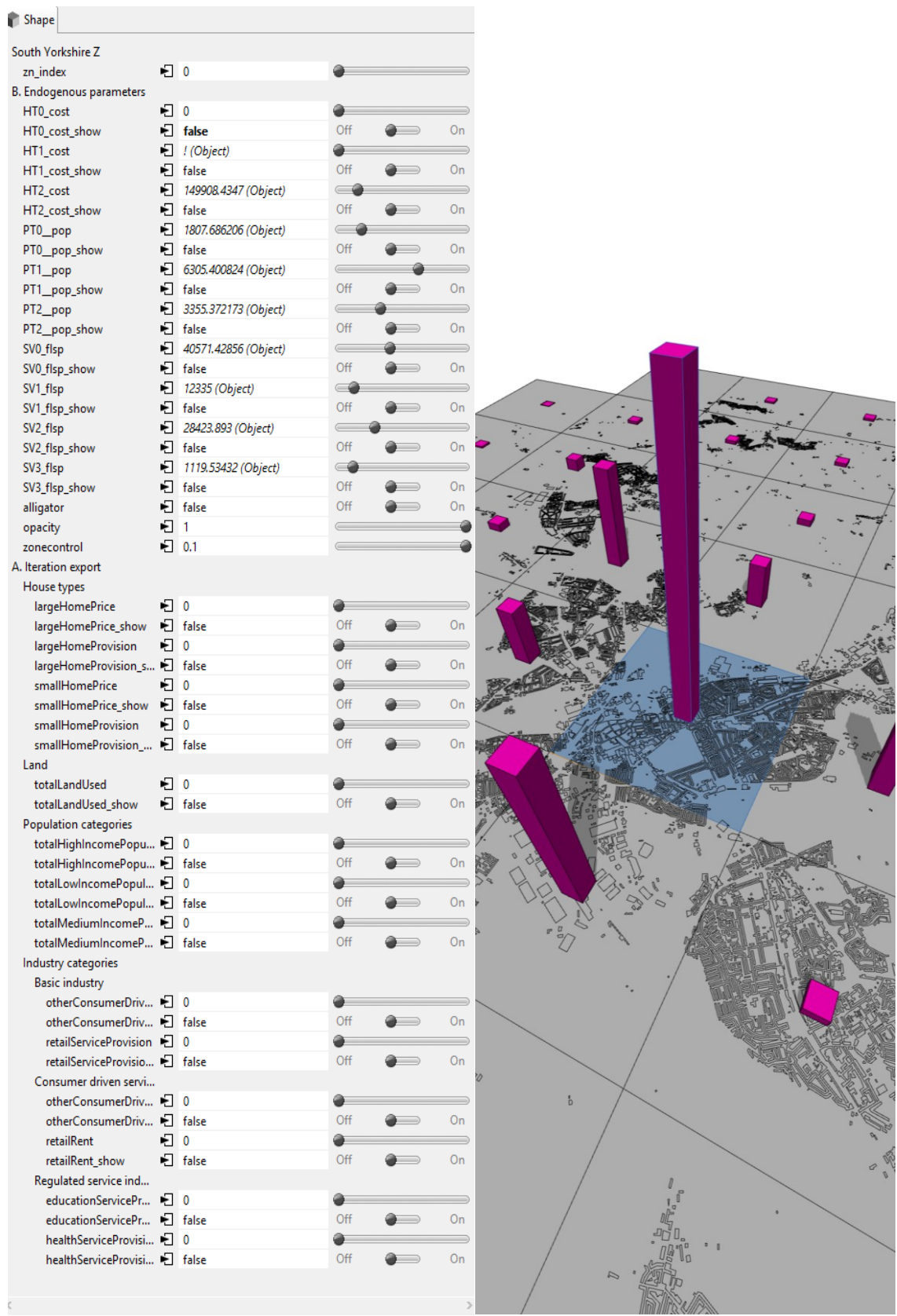

Figure 7. Custom CityEngine interface for the Dynamic Lowry Model. The controls expose exogenous parameters and allow users to adjust the parameters, toggle visualisation layers, compare model outputs, and inspect zonal attributes interactively.

## 4.4 Lower-tier implementation

For the lower tier we use Doncaster's Unitary Development Plan[3] to define planning zones (residential, industrial, services, education, green belt) and OS MasterMap[4] building footprints enriched with metadata such as heights land use and densities (Figure 8). The lower-tier model translates DLM indicators into local allocations and checks them against land-use capacity.

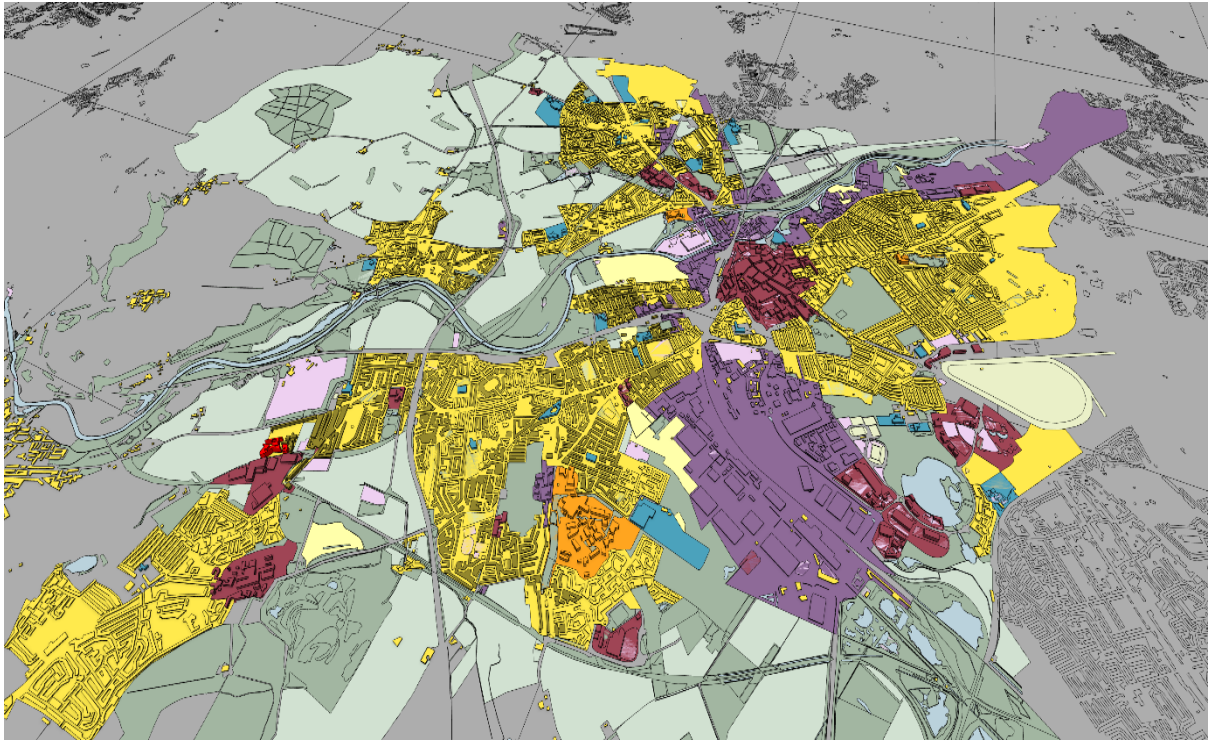

Figure 8. Doncaster Unitary Development Plan (2018) imported into CityEngine. Land-use categories are colour-coded (yellow = residential, red = retail, purple = industrial, blue = services, navy = education) to define the lower-tier planning zones used in the model.

As in case 1 we don't have detailed employment data for the lower tier, but we do have land uses, and population. We can then follow case 1 to allocate employment to the dedicated zone system in the masterplan.

This process involves three main steps:

1. Employment allocation:

$$e_i = \sum T_{ji} + T_{Ji} \quad \text{for } J \neq K \text{ and } i, j \in K \qquad (24)$$

Jobs are distributed to local zones based on both internal flows and inflows from external regions.

---

[3] The data in the maps contains public sector information licensed under the Open Government Licence v1.0. map available in http://doncaster.opo s3.co.uk/
[4] OS MasterMap Building Heights Layer [Shape geospatial data], Updated: Jul 2012, Ordnance Survey, Using: EDINA Digimap Ordnance Survey Service, Downloaded: September 2013.

2. Population allocation:

$$P_j = \sum T_{ij} + T_{Ij} \quad \text{for } I \neq K \text{ and } i, j \in K \qquad (25)$$

Similarly, Population flows are assigned to residential zones, ensuring regional growth translates into housing demand. Services are allocated in the same way.

3. Capacity checks using geometry:

CityEngine calculates the floorspace for each planning zone directly from geometry given by:

$$F_j = h_j \; H^f \; a_j \qquad (26)$$

where h is the height of the structure, $H^f$ the average floor height and *a* is the footprint area. Summing over all zones in K provides the aggregate land-use area requirement:.

$$(27)$$

$$A_K^R = \sum_{j \in K} A_j^R = \sum_{j \in K} F_j$$

Where $A^R_i$ is the total area of the zone j for sector R.

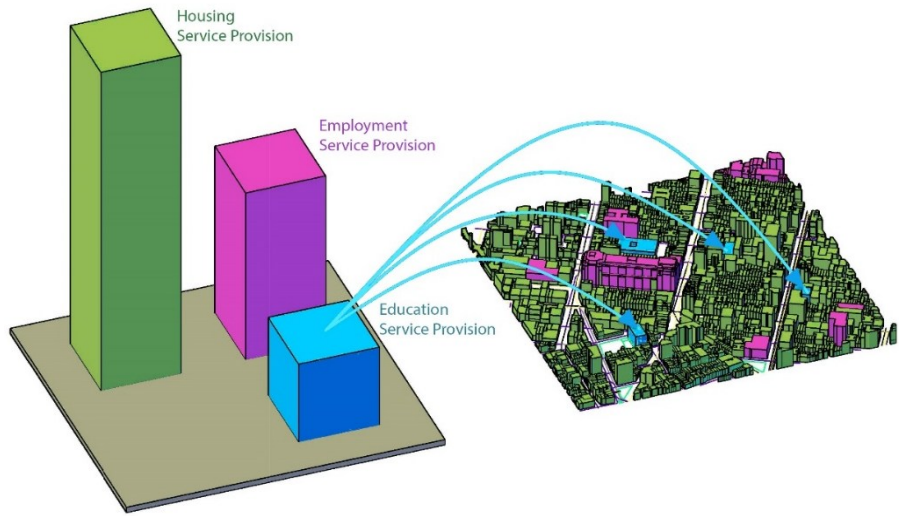

Figure 9. Linking regional indicators to planning zones. The figure demonstrates how DLM outputs such as education or employment provision are mapped to specific lower-tier zones for capacity evaluation.

To avoid exceeding what a zone can realistically accommodate, we enforce maximum capacities:

$$if \quad P_j > P_j^{max} \quad then \quad W_j \leftarrow W_j \cdot \frac{P_j^{max}}{P_j} \qquad (28)$$

Overloaded zones are highlighted, prompting the planner to redesign layouts or increase density.

Planner-facing controls (Figure 10) group variables by Planning Restrictions, Model Attributes, Land Use, and Geometry, enabling users to adjust assumptions, visualise impacts, and test scenarios interactively.

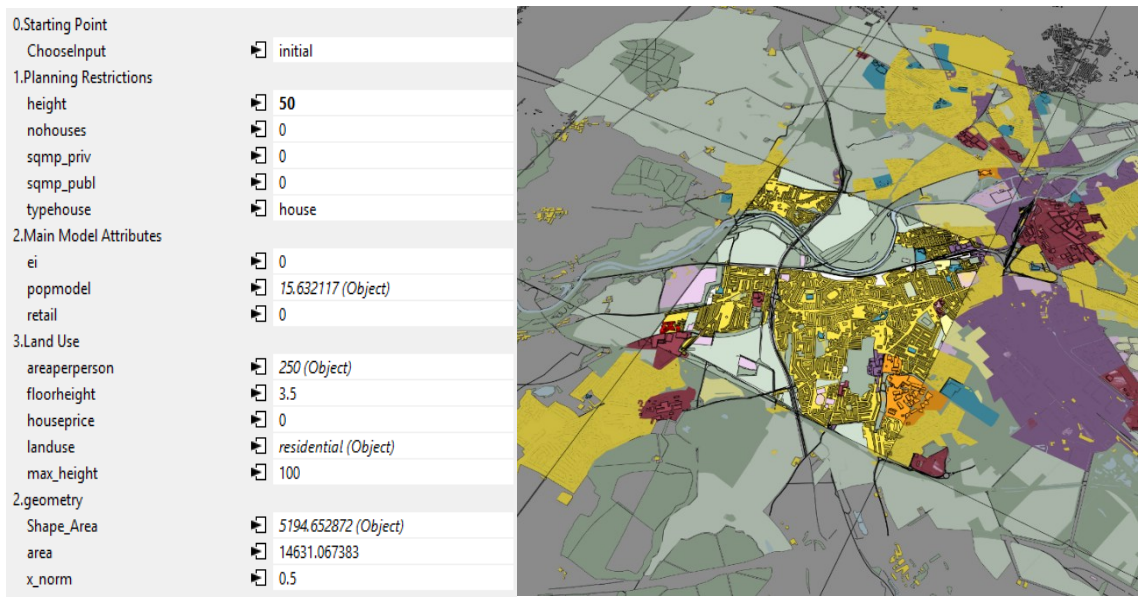

Figure 10. Planner-facing interface for the lower-tier model, showing the controls for land use, planning constraints, and geometry parameters in CityEngine, allowing users to adjust assumptions and visualise the impacts of different development scenarios in real time.

At the neighbourhood scale, the two-tier model works by taking regional forecasts (jobs, population, services) and distributing them into local planning zones. Employment flows are broken down to neighbourhoods, population flows translated into housing demand, and service demand checked against local capacity. CityEngine's geometry-based approach ensures that allocations are tied to physical form: if demand exceeds available floorspace, the system highlights the issue and prompts the planner to explore options such as adding new development zones, increasing densities, or reallocating land uses. In this way, regional change is directly connected to the physical design of neighbourhoods, and planners can test solutions interactively in 3D.

Figure 11 illustrates the full technical workflow used to implement the model in CityEngine. External data and planning constraints feed into Python scripts that allocate population and services between upper- and lower-tier zones using spatial interaction methods. These allocations are checked against land use capacity and adjusted iteratively. The results are linked to CGA visualization for geometry and user interaction, allowing the model to represent both exogenous inputs (e.g., employment factors, service demand, zoning rules) and endogenous outcomes (population distribution, floorspace provision). This integration enables scenario testing under different development and planning conditions.

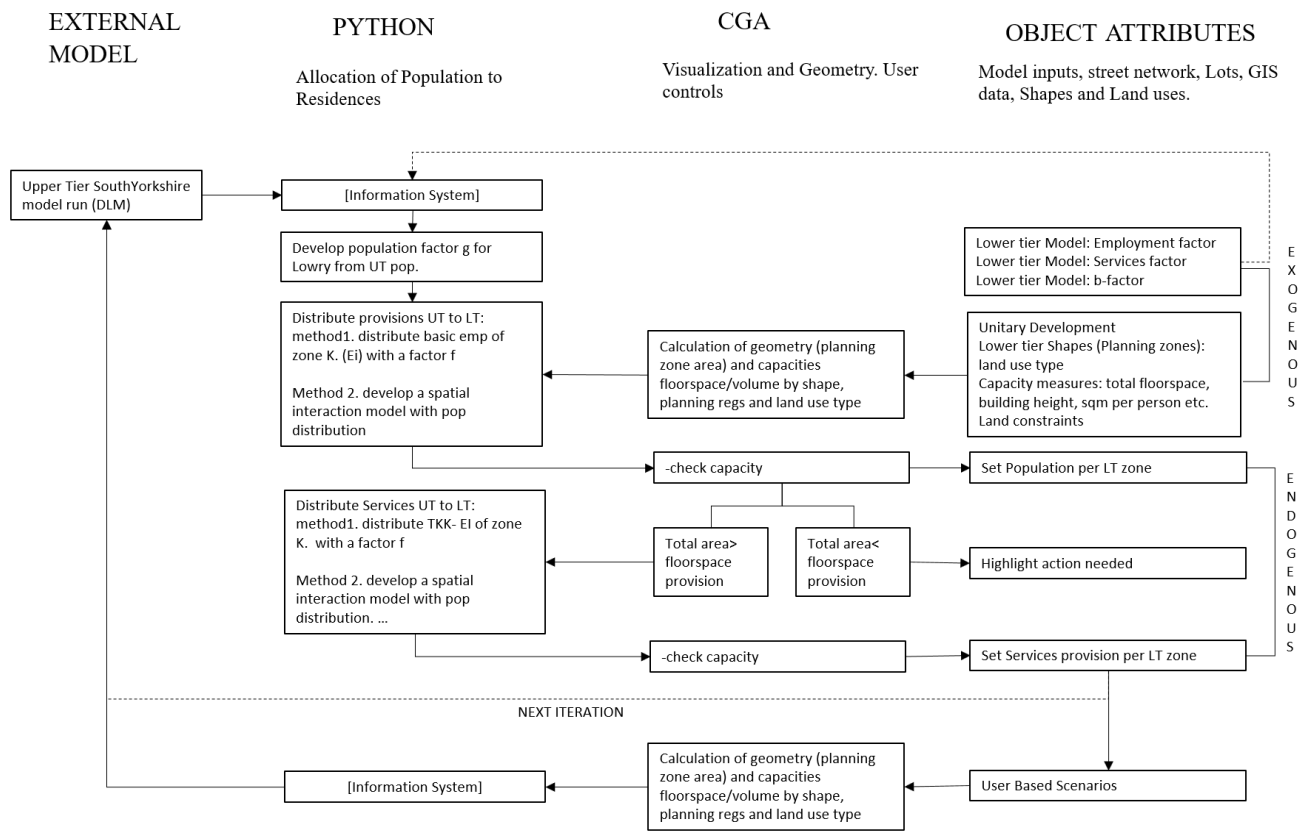

Figure 11. Framework of the two-tier CityEngine system. The diagram summarises how regional forecasts, local allocations, and design combine within a single environment to support multi-scale scenario planning.

The two-tier workflow is designed to be computationally efficient. At the regional level, the Dynamic Lowry Model (DLM) is fully calibrated and runs automatically, exporting year-by-year iterations almost instantaneously. These outputs are passed directly into CityEngine, where they can be visualised as an annual animation at around one frame per second, tagging indicators to the upper-tier zones in near real time. At the neighbourhood scale, the lower-tier allocation is somewhat heavier: distributing flows and tagging all planning zones for a single study area typically requires around 3–5 minutes. This balance ensures that regional dynamics can be simulated and streamed rapidly, while local allocations remain fast enough to support interactive design exploration.

In practice, this enables planners to create scenarios in two complementary ways:

1. By applying regional shocks (population or employment changes) in the DLM, which propagate down to neighbourhoods.

2. By modifying design, policies, land uses and capacities in CityEngine, which feed back into regional totals.

When a scenario exceeds capacity, it is flagged in real time, prompting planners to explore design-led alternatives such as new development zones, density increases, or land-use reallocation.

While the system demonstrates how regional forecasts can be translated into 3D planning scenarios, these lower-tier allocations are illustrative and not fully calibrated to real-world planning data. The purpose is to demonstrate methodology and technical implementation rather than deliver site-accurate forecasts, a distinction that is important before moving to applied scenarios in the next section.

## 5. Applications and Scenarios

To demonstrate how the two-tier model can be applied in practice, we present two simplified scenarios from South Yorkshire. These are not intended as full planning exercises, but as illustrative use cases that show how regional dynamics can be translated into local design tasks. The first scenario focuses on housing demand resulting from an employment shock, while the second addresses service provision by modelling school catchments. Together, they highlight the flexibility of the two-tier framework: it can respond both to changes in the regional economy and to the allocation of neighbourhood-scale social infrastructure.

**Scenario 1: Growth and housing demand.**

We first test a scenario of employment growth in South Yorkshire. Zone K (upper-tier zone 120) is used as the area of interest. Both the DLM (upper tier) and the lower-tier model are first run on baseline data to establish a calibrated reference.

We then simulate an increase in employment by adding over 1,000 new jobs in two neighbouring regional zones (zones 100 and 98), representing the development of new industrial or retail centres (Figure 12). The upper-tier run redistributes population accordingly, producing higher inflows into zone K. The lower-tier model translates these flows into housing demand, estimating that around 450 additional dwellings are required within zone K (Figure 13).

At this stage, the planner can respond in two ways:

- **Regional adjustment:** by modifying employment/residential provision values in the upper-tier system and re-running the DLM.

- **Local design response:** by editing land uses in CityEngine—for example, converting brownfield land into residential plots or laying out a new housing development.

The model then updates floorspace, capacity, and service demand in real time, highlighting any shortfalls (e.g., in retail or education provision). This scenario demonstrates how the two-tier framework can move seamlessly from **regional shocks** to **neighbourhood-scale design**, allowing planners to test different layouts and immediately see their implications.

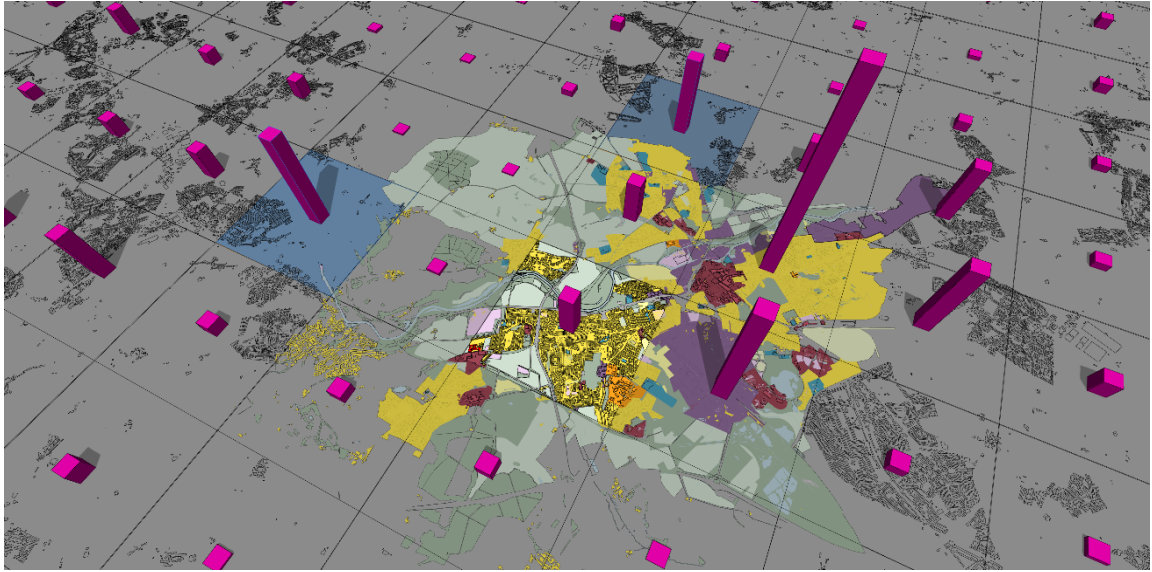

Figure 12. Regional employment shock scenario. The map illustrates an upper-tier simulation where new jobs are introduced in two regional zones (*I*).

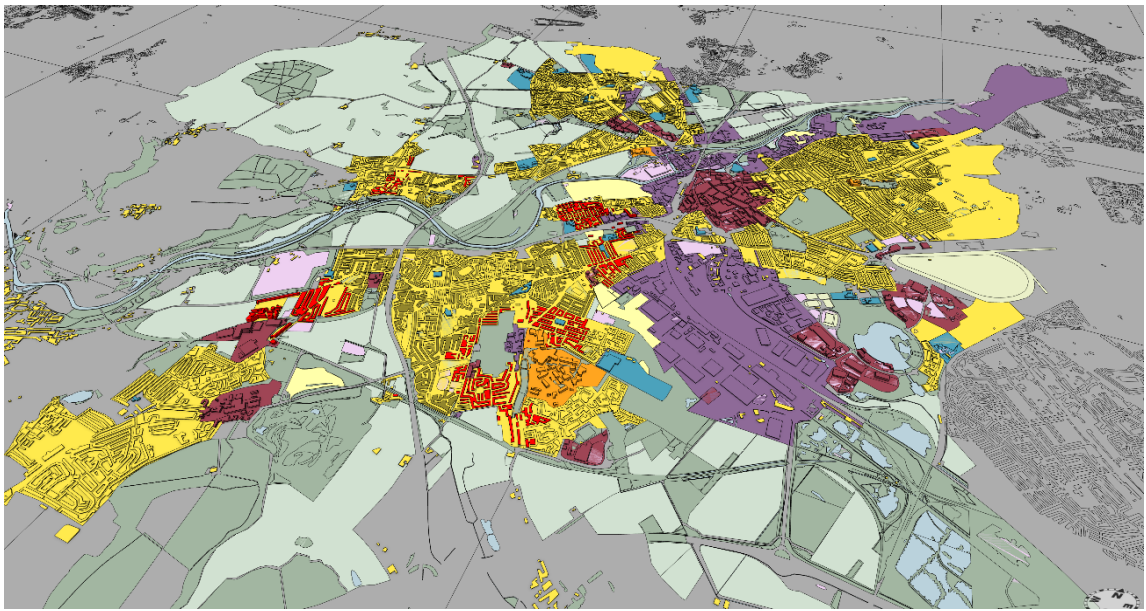

Figure 13. Housing demand in zone *K* following the employment growth scenario. The visualisation highlights areas of additional residential need (shown in red) resulting from new regional employment centres.

In addition to regional employment shocks, the two-tier system can also simulate changes within zone K itself. In the screenshots of Figure 14 and 15, we illustrate an example of a new development scenario, which is planned following the new industrial zone designed within zone K. The new housing development, covers most of the population demand in floorspace, however there is an additional demand in retail facilities. The development is designed within CityEngine, by changing the land use of brownfield land into residential using the lower-tier controls, and creating layout for building volumes and plots within this area.

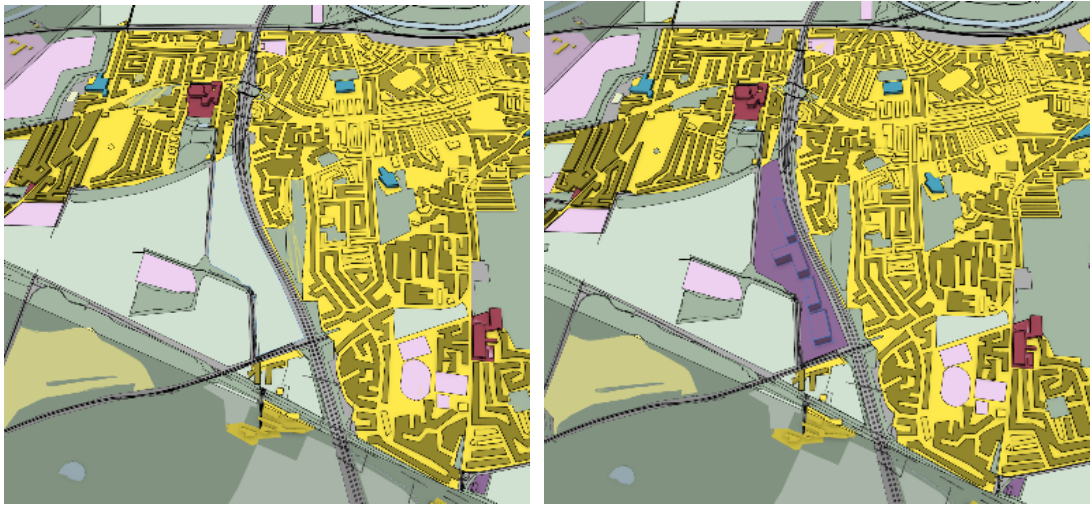

Figure 14. Allocation of a new industrial zone designed in CityEngine within zone *K*, located along the railway corridor.

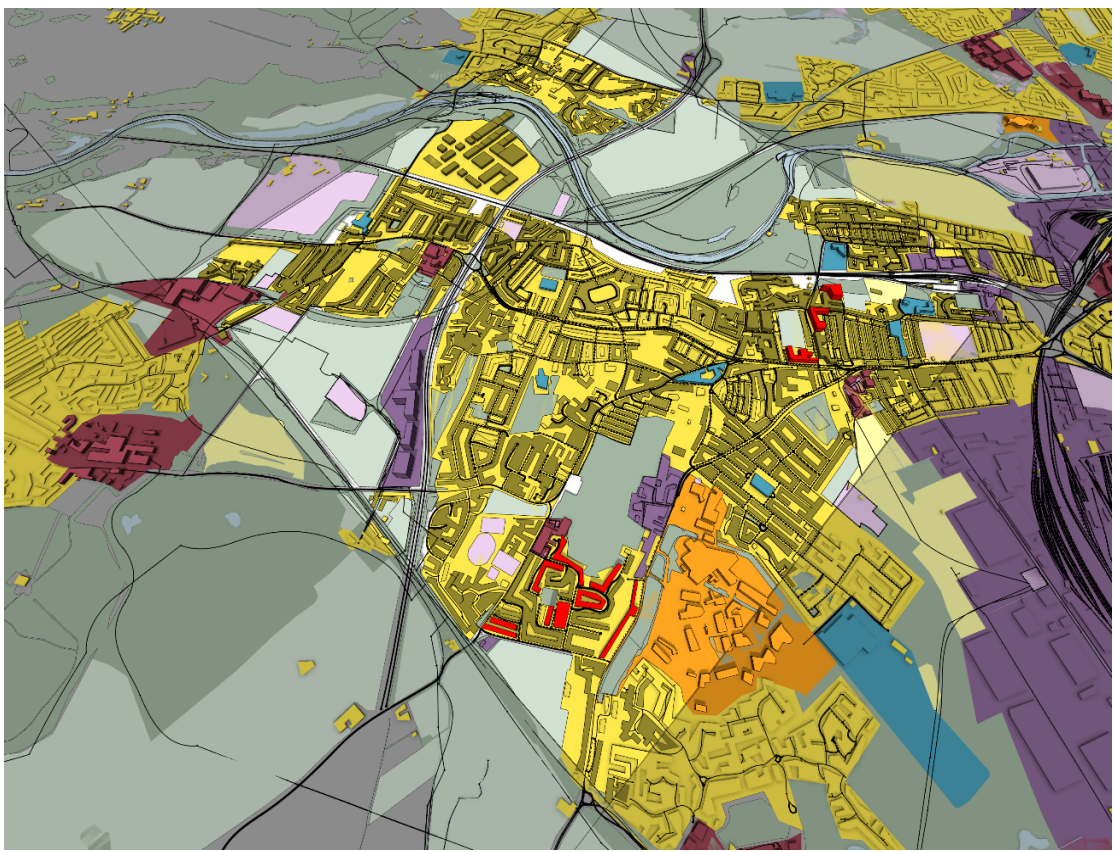

Figure 15. New housing demand resulting from the increased employment within zone K, illustrated with an example of a new residential development in the northern area. The model incorporates the added capacities into its allocation, updating floorspace and service requirements accordingly.

**Scenario 2: New schools.**

A second example focuses on the allocation of schools following population growth. Unlike retail or housing, schools are regulated services that must be planned according to catchment size and capacity. In standard practice, buffer zones are drawn around each school type (primary, secondary, etc.), but these often fail to account for population distribution, accessibility, or changing demand.

Using the two-tier model, we expand zone K to include adjacent areas (Case 3) and assign residential zones as origins and schools as destinations. Each school is given a maximum capacity, and flows of students are simulated from neighbourhoods to schools. This produces catchment areas based on flows rather than buffers, showing how many pupils each school will serve (Figure 16).

When a new employment or population scenario is introduced in the upper tier, the system recalculates demand for schools and highlights capacity gaps. Planners can then test the effects of siting a new school in different locations within zone K and immediately see how catchments, enrolments, and service provision change. This scenario demonstrates how the two-tier framework can be used to plan social infrastructure, linking demographic shifts with service allocation in an interactive 3D environment.

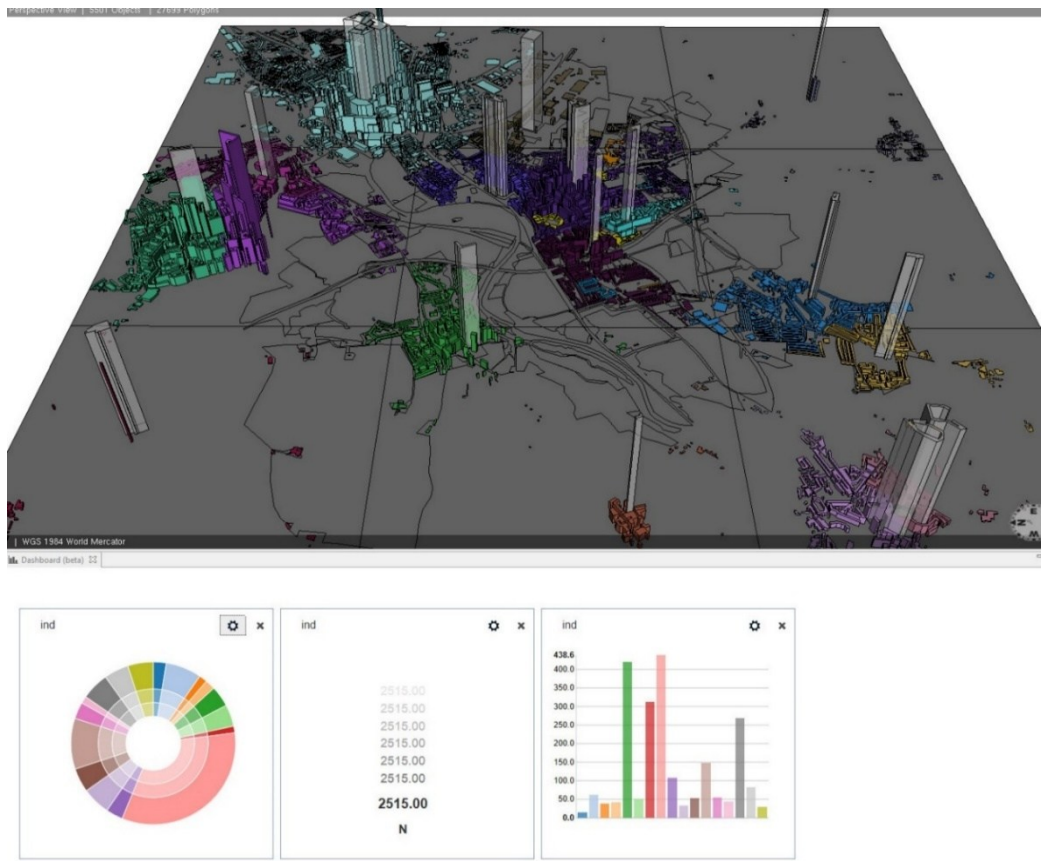

Figure 16. Modelled school catchment areas for an expanded *K* zone and Analytics. The flows-based approach defines catchments by student assignments rather than fixed buffers, allowing planners to assess service capacity and potential gaps.

## 6. Limitations

Despite its advantages, the two-tier approach has limitations. The model currently relies on simplified cost functions (Euclidean distance rather than travel times) and does not yet incorporate uncertainties or sensitivity analyses. The lower-tier implementation is illustrative rather than calibrated to detailed planning data, and the CityEngine workflow requires manual input to test scenarios rather than providing automated optimisation. Moreover, the case study focused on South Yorkshire, and while the modular structure is transferable, its generality has not been tested in other contexts. These limitations point directly to avenues for further work.

## 7. Conclusions and future work

This paper has demonstrated how a two-tier Lowry model can link regional forecasts with neighbourhood-scale planning. By introducing regional flows into the design of local masterplans, the framework ensures that development scenarios account for external dependencies such as commuting, housing demand, and service provision. The South Yorkshire case study illustrated how outputs from the Dynamic Lowry Model can be coupled with the Doncaster Unitary Development Plan to translate regional change into local housing and service needs. The main contribution of this work is a modular workflow that connects modelling, planning, and visualisation tasks while keeping them distinct, allowing different models or datasets to be swapped in without disrupting the system. Implemented in CityEngine, the approach also leverages procedural rules to visualise multiple scenarios quickly and to highlight capacity constraints.

The workflow is computationally efficient. At the regional level, the Dynamic Lowry Model runs automatically and exports annual forecasts almost instantaneously. These outputs are streamed directly into CityEngine, where they can be animated at around one frame per second to track change year by year. At the neighbourhood scale, lower-tier allocations require on the order of 2–5 minutes per zone depending on number of zones, a speed sufficient to support interactive planning and design exploration. This balance ensures that the framework is not only theoretically rigorous but also practical for day-to-day use.

Although this paper has demonstrated the method with a Dynamic Lowry Model, the framework is modular in principle: any regional model, such as QUANT or other LUTI systems, could serve as the upper tier, provided that zonal forecasts and indicators are exported and tagged appropriately. This flexibility positions the two-tier approach as a bridge rather than a competitor, complementing existing models by translating their outputs into tangible, testable 3D design scenarios. In this sense, the work responds directly to Couclelis' (1997) call for urban models that are not only analytically robust but also accessible and usable as planning support systems.

Future work should focus on extending the model to include accessibility-based cost functions, richer planning indicators (e.g., energy use, emissions, equity), and more detailed architectural typologies in the visualisation layer. Integrating live data feeds from regional models would move the system toward a dynamic, real-time planning tool, while testing it in other cities and with alternative regional models (e.g., QUANT, Mechanicity) would strengthen its transferability. Taken together, these steps would advance the two-tier approach as a practical framework for bridging the divide between forecasting models and urban design practice.

## 8. Acknowledgements

This research originated as part of the author's doctoral work at the Bartlett Centre for Advanced Spatial Analysis (CASA), University College London, supervised by prof. Andrew Hudson-Smith and Sir Alan Wilson. It was supported by the UK Engineering and Physical Sciences Research Council (EPSRC).